\documentstyle[11pt,moriond,epsfig]{article}
\newcommand{\nc}{\newcommand}
\nc{\hc}{\hbox {h.c.}}
\nc{\re}{\hbox {Re}}
\nc{\im}{\hbox {Im}}
\nc{\etal}{\hbox{et al.}}
\nc{\ra} {\rightarrow}
\nc{\cw}{\cos\theta_W}        \nc{\sw}{\sin\theta_W}
\nc{\ttbar}{t\bar{t}}
\nc{\bbbar}{b\bar{b}}
\nc{\tanb} {\tan \beta}
\nc{\twbdec} {t\rightarrow W^+ b}
\nc{\tbwbdec} {\bar{t} \rightarrow W^- \bar{b}}
\nc{\hprod} {e^+e^- \ra Z^\ast \ra H Z}
\nc{\epem} {e^+e^-}
\nc{\wpwm} {W^+W^-}
\nc{\tbar} {\bar{t}}
\nc{\bbar} {\bar{b}}
\nc{\wpp} {W^+}
\nc{\mt}{m_t}
\nc{\mts}{m_t^2}
\nc{\mw} {m_W}
\nc{\mws} {m_W^2}
\nc{\mz} {m_Z}
\nc{\mzs} {m_Z^2}
\nc{\mh} {m_H}
\nc{\mhs} {m_H^2}
\nc{\ma} {m_A}
\nc{\mas} {m_A^2}
\nc{\hdec}{H \ra t\bar{t}}
\nc{\ttbardec}{\ttbar \ra W^+W^-\bbbar}
\nc{\po}{\Phi_1}
\nc{\pod}{\Phi_1^\dagger}
\nc{\pht}{\Phi_2}
\nc{\phtd}{\Phi_2^\dagger}
\nc{\phtt}{{\tilde{\Phi}}_2}
\nc{\popo}{\po^\dagger\po}
\nc{\phtpt}{\pht^\dagger\pht}
\nc{\popt}{\po^\dagger\pht}
\nc{\phtpo}{\pht^\dagger\po}
\nc{\sq}{\sqrt{2}}
\nc{\nsd} {N_{SD}}
\nc{\ntt} {N_{tt}}
\nc{\vs}{\vspace{2mm}}
\nc{\sty}{\hat{S}^t_1} \nc{\pty}{\hat{P}^t_1}
\nc{\sts}{(\sty)^2}      \nc{\pts}{(\pty)^2}
\nc{\yts}{\sts+\pts}
\nc{\sby}{\hat{S}^b_1} \nc{\pby}{\hat{P}^b_1}
\nc{\sbs}{(\sby)^2}      \nc{\pbs}{(\pby)^2}
\nc{\ybs}{\sbs+\pbs}

\def\lsim{\mathrel{\raise.3ex\hbox{$<$\kern-.75em\lower1ex\hbox{$\sim$}}}}
\def\gsim{\mathrel{\raise.3ex\hbox{$>$\kern-.75em\lower1ex\hbox{$\sim$}}}}

\def\ra{\rightarrow}

\def\be{\begin{equation}}
\def\ee{\end{equation}}
\def\bea{\begin{eqnarray}}
\def\eea{\end{eqnarray}}

\begin{document}

\hfill IFT-99/26

\hfill KIAS-P99100

\hfill October 1999

\vspace*{4cm}
\title{RECONSTRUCTION OF BASIC PARAMETERS FROM CHARGINO 
PRODUCTION \footnote{Talk given by J. Kalinowski 
at the XIth Recontres de Blois, ``Frontiers of Matter'', 
June 27-July 3, 1999, France}}

\author{ S.Y. CHOI$^1$ and  J. KALINOWSKI$^{2}$ }

\address{$^1$ Korea Institute for Advanced Study, Seoul 130-012, Korea  \\
         $^2$ Instytut Fizyki Teoretycznej UW, 
Ho\.za 69, 00681 Warsaw, Poland }

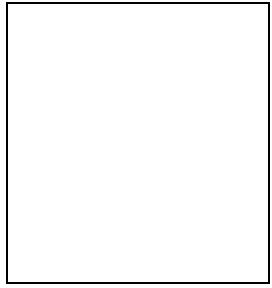
\maketitle\abstracts{The event characteristics of chargino pair
production at $e^+e^-$ collisions are explored to determine
gaugino-higgsino mixing angles. We demonstrate that by measuring total
cross sections and left-right asymmetries with polarized beams or
angular correlations among chargino decay products the fundamental
SUSY parameters $M_2$, $\mu$ and $\tan\beta$ can be uniquely
determined in a model independent way. }

\newpage
\section{Introduction}
The low-energy supersymmetric model (SUSY) involves a large
number of arbitrary parameters reflecting our ignorance of the
supersymmetry breaking mechanisms.  Once the supersymmetric particles
are discovered, the priority will be to measure the low-energy SUSY
parameters independently of theoretical assumptions and confront them
with relations following from e.g. grand unification theories
\cite{lc}.  A clear strategy is needed to deal with so many a priori
arbitrary parameters \cite{epi99}.

In many scenarios charginos, mixtures of spin 1/2 partners of the $W$
and charged Higgs bosons, $\tilde{W}^\pm$ and $\tilde{H}^\pm$, 
are among the lightest supersymmetric
particles. The chargino sector depends on only three soft SUSY
breaking parameters: the SU(2) gaugino mass $M_2$, the higgsino mass
parameter $\mu$, and the ratio $\tan\beta(=v_2/v_1)$ of the vacuum
expectation values of the two neutral Higgs fields 
which enter the chargino mass matrix 
($s_{\beta}=\sin\beta$, $c_{\beta}=\cos\beta$)
\begin{eqnarray}
{\cal M}_C=\left(\begin{array}{cc}
                M_2                &      \sqrt{2}m_W c_\beta  \\
             \sqrt{2}m_W s_\beta  &             \mu   
                  \end{array}\right)
\label{eq:mass matrix}
\end{eqnarray}
written in the ($\tilde{W}$, $\tilde{H}$) basis \footnote{In 
CP--noninvariant theories,  
$M_2$ and  $\mu$ can be complex.  However,
by reparametrization of the fields, $M_2$ can be assumed real and
positive without loss of generality so that only 
$\mu=|\mu|\exp i \Phi_{\mu}$ may have a non--trivial
invariant phase $\Phi_{\mu}$. 
Here we will consider a CP--invariant scenario.}. 
Therefore the chargino production processes 
\begin{eqnarray*}
e^+ e^- \ \rightarrow \ \tilde{\chi}^-_i \ \tilde{\chi}^+_j \; 
\; \; \; \; \; [i,j=1,2]
\,.
\end{eqnarray*}
may serve as a good starting point towards a systematic and 
model--independent
determination of the fundamental SUSY  parameters \cite{choi1,choi2}.

\section{Exploiting chargino production processes}
Since the mass matrix (\ref{eq:mass matrix})
is asymmetric, two different mixing matrices acting on the left-- and
right--chiral $(\tilde{W},\tilde{H})$ states are needed to diagonalize
it.  The two eigenvalues are given by
\begin{eqnarray}
m^2_{\tilde{\chi}^\pm_{1,2}}
  ={\textstyle \frac{1}{2}}[M^2_2+\mu^2+2m^2_W
    \mp \Delta \, ]
\end{eqnarray}
and the left-- and right--chiral components of the light mass eigenstate 
$\tilde{\chi}^-_1$ are related to the wino and higgsino components
in the following way  
\begin{eqnarray}
\tilde{\chi}^-_{1L}=\tilde{W}^-_L\cos\phi_L
                     +\tilde{H}^-_{1L}\sin\phi_L, \;\;\;\; 
\tilde{\chi}^-_{1R}=\tilde{W}^-_R\cos\phi_R
                     +\tilde{H}^-_{2R}\sin\phi_R, 
\end{eqnarray}
%
\begin{eqnarray}
&&\cos 2\phi_L=-(M_2^2-\mu^2-2m^2_W\cos 2\beta)/\Delta, \;\;\; 
\sin 2\phi_L=-2\sqrt{2}m_W(M_2\cos\beta+\mu\sin\beta)/\Delta
\nonumber\\
&&\cos 2\phi_R=-(M_2^2-\mu^2+2m^2_W\cos 2\beta)/\Delta, \;\;\; 
\sin 2\phi_R=-2\sqrt{2}m_W(M_2\sin\beta+\mu\cos\beta)/\Delta\nonumber \\
&&{}
\rule{3cm}{0cm}
\Delta=[(M^2_2+\mu^2+2m^2_W)^2-4(M_2\mu-m^2_W\sin 2\beta)^2]^{1/2}
\label{mixing}
\end{eqnarray}

\subsection{Production of light charginos}
With only light charginos $\tilde{\chi}^{\pm}_1$
accessible kinematically, the parameters $M_2,\mu$ and
$\tan\beta$ can be determined up to at most a two-fold discrete
ambiguity \cite{choi1}.  To
this end measurements of the chargino mass, the total production cross
section and angular correlations among the chargino decay products are
necessary. The angular correlations depend on the polarizations
of the produced charginos and thus on the
gaugino--higgsino mixing angles $\phi_L$ and $\phi_R$. Beam
polarization is helpful but not necessarily required.  

Since charginos decay through 
a $W$ boson  or scalar partners of leptons or quarks, 
the decay matrix
depends on additional parameters, like  scalar masses and
couplings to neutralinos. The presence of two
invisible neutralinos in the process $ e^+ e^- \to
\tilde{\chi}_1^+ \tilde{\chi}_1^- \to \tilde{\chi}_1^0 
\tilde{\chi}_1^0 (f_1 \bar{f}_2) (\bar{f}_3
f_4)$ makes it impossible to measure directly the chargino production
angle $\Theta$. Integrating over this angle
and also over the invariant masses of the fermionic systems $(f_1
\bar{f}_2)$ and $(\bar{f}_3 f_4)$, the fully correlated angular distribution
$\Sigma(\theta^*, \phi^*,  \bar{\theta}^*, \bar{\phi}^*)$ 
can be expressed in terms of sixteen independent 
angular combinations of helicity production  amplitudes.
The ($\theta^*$,$\phi^*$)  are the polar and azimuthal angles
of the ($f_1 \bar{f}_2$) system in the $\tilde{\chi}^-_1$ rest frame with
respect to the chargino's flight direction in the lab frame;
quantities with a bar refer to the $\tilde{\chi}^+_1$ decay. 
Out of the sixteen terms, corresponding to the unpolarized,
$2 \times 3$ polarization components and $3 \times 3$ spin--spin
correlations, only 7 are independent 
if  small effects from the $Z$-boson width
and loop corrections are neglected.

The crucial observation of \cite{choi1} is that four terms:
$\Sigma_{un}, {\cal P}, {\cal Q}$ and ${\cal Y}$, can be measured 
{\it directly} 
by means of simple kinematical projections since $\cos \theta^*, \cos
\bar{\theta^*}$ and $\sin \theta^* \sin \bar{\theta^*} \cos( \phi^* +
\bar{\phi^*})$ are simple functions of  
energies and momenta of the decay systems
($f_i \bar{f_j}$) in the lab frame. Moreover, three physical
observables, $\Sigma_{un}, {\cal P}^2/{\cal Q}$ and ${\cal
P}^2/{\cal Y}$ by construction are independent of the details of the
chargino decay dynamics and of the structure of (potentially more complex)
neutralino and sfermion sectors.
The measurements of $\sigma_{tot}$ and either of the ratios
${\cal P}^2/{\cal Q}$ or ${\cal P}^2/{\cal Y}$ therefore allow us to
determine $\cos 2\phi_L$ and $\cos 2\phi_R$. 
If polarized beams are available, the
left-right asymmetry $A_{LR}$ can provide an alternative way to
extract these quantities (or serve as a consistency check). From the
``measured'' values of $\{\cos 2\phi_L,\cos 2\phi_R\}$ and
the chargino mass, the  Lagrangian parameters $M_2$, $\mu$
and $\tan\beta$ can be obtained  up to a two-fold ambiguity \cite{choi1}.
\subsection{Above the heavy chargino threshold}
It has been recently demonstrated \cite{choi2} that if the collider
energy is sufficient to produce the light and heavy chargino states in 
pairs, the
underlying fundamental SUSY parameters, $M_2,\mu$ and $\tan\beta$, can
be extracted {\it unambiguously} from chargino masses, 
production cross sections  and
left-right asymmetries with polarized electron beams. The new
ingredient is the heavier chargino mass which, like for the lighter
chargino, can be determined very precisely from the sharp rise of the
cross sections
$\sigma(e^+e^-\rightarrow\tilde{\chi}^-_i\tilde{\chi}^+_j)$.
The value of $\tan\beta$ is uniquely determined in terms of the
mass difference of two chargino states, $\Delta =
m^2_{\tilde{\chi}^\pm_2}-m^2_{\tilde{\chi}^\pm_1}$, and two mixing
angles as follows
\begin{eqnarray}
\tan\beta=[(4m^2_W
                     +\Delta \,
                      (\cos 2\phi_R-\cos 2\phi_L))/(4m^2_W
                     -\Delta \,
                      (\cos 2\phi_R-\cos 2\phi_L))]^{1/2}
\label{eq:tanb}
\end{eqnarray}
and  $M_2$, $|\mu|$ and $\mbox{sign}(\mu)$ \footnote{In 
the CP--noninvariant theories, the
eq.~(\ref{eq:signmu}) determines the $\cos\Phi_{\mu}$.}  
    are given by
\begin{eqnarray}
 M_2&=&{\textstyle \frac{1}{2}}
        [2(m^2_{\tilde{\chi}^\pm_2}+m^2_{\tilde{\chi}^\pm_1}-2m^2_W)
              -\Delta \, 
               (\cos 2\phi_R+\cos 2\phi_L)]^{1/2}\nonumber\\
 |\mu|&=&{\textstyle \frac{1}{2}}
        [2(m^2_{\tilde{\chi}^\pm_2}+m^2_{\tilde{\chi}^\pm_1}-2m^2_W)
              +\Delta \, 
               (\cos 2\phi_R+\cos 2\phi_L)]^{1/2} 
\label{eq:M2mu}\\
\mbox{sign}(\mu)&= &[ \Delta^2
                   -(M^2_2-\mu^2)^2-4m^2_W(M^2_2+\mu^2)
                   -4m^4_W\cos^2 2\beta]/8 m_W^2M_2|\mu|\sin2\beta
\label{eq:signmu}
\end{eqnarray}
\subsection{Expected errors}
An important question is what precision one might expect in the above
procedure due to the propagation of experimental errors.  It turns out
that  $M_2$ and $|\mu|$ can be determined
very well from eqs.~(\ref{eq:M2mu}).  However for large $\tan\beta$ case, the
eq.~(\ref{eq:tanb}) is not very useful to obtain the value of
$\tan\beta$ due to error propagation. Instead, we can determine first
$\cos2\beta=\Delta (\cos 2\phi_L-\cos 2\phi_R)/4 m_W^2$
and check for the sign of $\mu$ from the numerator of
eq.~(\ref{eq:signmu}). For this purpose the error of $\cos2\beta$ does
not matter. Then the eq.~(\ref{eq:signmu}) can be used to extract
$\sin2\beta$ and finally $\tan\beta$ obtained.  As an example we
consider two reference points in the parameter space, denoted by RR1
and RR2 in \cite{choi2}, corresponding to $(M_2,\mu,\tan\beta)=(152, \,
316, \, 3)$ and $(150,\, 263, \, 30)$, respectively (mass parameters
are in GeV). We consider a high luminosity option of 500 fb$^{-1}$
at the $e^+e^-$ c.m. energy $\sqrt{s}=800$ GeV and 
only statistical errors are taken into account. 
From measured total cross sections and LR asymmetries  
the mixing angles can be determined with the  
expected errors as follows:
$\delta\cos2\phi_L=0.02$ and $\delta\cos2\phi_R=0.005$. 
Assuming the error of 100 MeV for the chargino masses,  
we arrive at the values shown in the Table:
\begin{center}
\begin{tabular}{c|c|c}
& RR1 & RR2\\
\hline
$M_2$ [GeV]    & $152 \pm 1.75$         & $150 \pm 1.2$ \\
$\mu$ [GeV]    & $316 \pm 0.85$         & 263 $\pm$ 0.68 \\
$\cos2\beta$   & $-0.8\pm 0.08$         & $ -0.998 \pm 0.056$\\
$\sin2\beta$   & $0.6 \pm 0.058$        & $ 0.066 \pm 0.033$
\end{tabular}
\end{center}

\noindent It is important to stress that for a discrimination between
small and large  $\tan\beta$ scenarios a high luminosity option is
required. From the numbers for $\sin2\beta$ one 
finds the ``measured'' values of 
$\tan\beta$ in the range $\{2.46,\, 4.01\}$  and $\{20.2,\, 59.6\}$ for 
RR1 and RR2, respectively. 

\section{Summary}

We explored event characteristics to isolate the chargino sector 
instead of relying on global fits to chargino/neutralino system \cite{mp}.  
The angular correlations among chargino decay
products provide two independent observables.  
Thus,   even from the light chargino pair production in $e^+e^-$
annihilation,  $\tan\beta$, $M_2$
and $\mu$ are determined up to at most a two-fold discrete
ambiguity.  
If the collider energy is sufficient to produce the two chargino
states in pairs, the above ambiguity is removed. 
The production cross sections and LR asymmetries allow us
to determine chargino masses and mixing angles very precisely 
and extract \footnote{Some of the material presented here
goes through unaltered if the phase $\Phi_{\mu}$ is 
allowed \cite{choi1,choi2}, although extra information is  
needed to determine the phase.}
 $\tan\beta$, $M_2$ and $\mu$ unambiguously. 

Note that from the energy distribution of chargino decay products 
the mass of the lightest neutralino can be 
measured. As a result, the parameter $M_1$  and 
the neutralino mass matrix can be also reconstructed \cite{choi2}. 
An alternative way 
to determine $M_1$, $M_2$, $\tan\beta$ and $\mu$, 
based only on some of the 
masses of charginos and neutralinos, can be
found in \cite{km}.

In short,  $e^+e^-\rightarrow \tilde{\chi}^-_i \
\tilde{\chi}^+_j$ with polarized beams allows us to extract the
fundamental parameters in the chargino sector.

\section*{Acknowledgments}
We acknowledge invaluable discussions with P.M. Zerwas. 
This work has been supported  by the KOSEF-DFG 
Project 96-0702-01-01-2 (SYC) and  the KBN Grant
  2~P03B~030~14 (JK).

\end{document}